\documentclass{emulateapj}

\begin{document}

\slugcomment{ApJ accepted}

\title{Ultra-Compact Dwarfs in the Core of the Coma Cluster
\altaffilmark{1}}


\author{ Juan P. Madrid\altaffilmark{2}, 
Alister W. Graham\altaffilmark{2},
William E. Harris\altaffilmark{3},\\
Paul Goudfrooij\altaffilmark{4},
Duncan A. Forbes\altaffilmark{2},
David Carter\altaffilmark{5},
John P. Blakeslee\altaffilmark{6},\\
Lee R. Spitler\altaffilmark{2},
Henry C. Ferguson\altaffilmark{4}}

\altaffiltext{1}{Based on  observations made with  the NASA/ESA Hubble
Space Telescope,  obtained at  the Space Telescope  Science Institute,
which is operated  by the Association of Universities  for Research in
Astronomy, Inc., under NASA  contract NAS5-26555.  These observations
are associated with program 10861.}

\altaffiltext{2}{Centre for Astrophysics and Supercomputing, Swinburne
University of Technology, Hawthorn, VIC 3122, Australia}

\altaffiltext{3}{Department of Physics and Astronomy, McMaster University,
Hamilton, ON L8S 4M1, Canada} 

\altaffiltext{4}{Space Telescope Science Institute, 3700 San Martin
drive, Baltimore, MD 21218, USA} 

\altaffiltext{5}{Astrophysics Research Institute, Liverpool John Moores University, Egerton Wharf, Birkenhead, UK} 

\altaffiltext{6}{Herzberg Institute of Astrophysics, Victoria, BC V9E 2E7, Canada} 



\begin{abstract}

We have discovered both a red and a blue subpopulation of Ultra-Compact Dwarf  (UCD) galaxy candidates in the Coma galaxy cluster. We analyzed deep F475W (Sloan $g$) and F814W (I) Hubble 
Space Telescope images obtained with the Advanced Camera for Surveys Wide Field Channel as part of the Coma Cluster Treasury Survey and have  fitted the light profiles of $\sim$5000 point-like 
sources in the vicinity of NGC 4874 --- one of the two central dominant galaxies of the Coma cluster. Although almost all of these sources are globular clusters that remain unresolved, we found that 52 
objects have effective radii between $\sim$10 and 66 pc, in the range spanned by Dwarf Globular Transition Objects (DGTO) and UCDs. Of these 52 compact objects, 25 are brighter than $M_V \sim 
-11$ mag, a magnitude conventionally thought to separate UCDs and globular clusters. The UCD/DGTO candidates have the same color and luminosity distribution as the most luminous globular 
clusters within the red and blue subpopulations of the immensely rich NGC 4874 globular cluster system. Unlike standard globular clusters, blue and red UCD/DGTO subpopulations have the same 
median effective radius. The spatial distribution of UCD/DGTO candidates reveal that they congregate towards NGC 4874, and are not uniformly distributed. We find a relative deficit of UCD/DGTOs compared with globular clusters in the inner 15 kpc around NGC 4874, however at larger radii UCD/DGTO and globular clusters follow the same spatial distribution.

\end{abstract}





\section{Introduction}

Within galaxy clusters, cD galaxies are swarming with tens of thousands of globular clusters. Like most galaxies 
with rich  Globular Cluster Systems (GCS), the cD galaxies of Virgo (M87: Peng et al.\ 2009; Harris 2009b) and Fornax (NGC 1399: Dirsch et al.\ 2003) display 
evident bimodality in their color distribution. Bimodal fits to the GCS 
around NGC 3311, the central cD galaxy of the Hydra galaxy cluster, also reveal the presence of a red (metal-rich) and a blue (metal-poor) subpopulation
(Wehner et al.\ 2008). Six additional giant elliptical galaxies studied 
by Harris (2009a) also show clear bimodality.

In the analysis of the color-magnitude diagram (CMD) of the NGC 3311 GCS,
Wehner \& Harris (2007; their Figure 1) find an "upward" extension towards 
brighter magnitudes of the red subpopulation. This continuation to brighter magnitudes (and higher masses) is absent among 
the blue subpopulation. Wehner \& Harris (2007) postulate that given 
their magnitudes ($i' \leq 22.5$ mag) and inferred masses ($>6\times10^6 M_{\odot}$), the extension of the CMD corresponds to Ultra-Compact Dwarfs (UCDs, Phillipps et al.\ 2001).  While Wehner \& 
Harris (2007) suggest that UCDs are the bright extension of the GCS, the ground-based data used for their work do not allow them to derive characteristic radii for their
UCD candidates. Peng et al.\ (2009) over-plotted 18 objects with extended effective radii ($r_{h} > 10$ pc) on their CMD of the M87 GCS. Some of these objects are confirmed UCDs or Dwarf/Globular 
Transition Objects (Hasegan et al.\ 2005) and most of them lie exactly at the bright end of the blue subpopulation of globular clusters.

Ultra-compact dwarfs are a relatively newly discovered class of stellar 
system (Hilker et al.\ 1999; Drinkwater et al. 2000) and their origin is not yet clearly established. In color, and structural parameters, UCDs lie between supermassive globular clusters and compact 
dwarf elliptical galaxies. Not surprisingly, the origin of UCDs has been linked to both globular clusters and dwarf elliptical galaxies (Hilker 2006, and references therein). UCDs are typically associated 
with galaxy clusters and were originally identified in the Fornax cluster (Hilker et al.\ 1999; Drinkwater et al.\ 2000, 2003). Subsequently, the presence of UCDs was reported in several additional 
galaxy clusters: Virgo, Centaurus, Hydra, Abell S040 and Abell 1689 (Evstigneeva et al.\ 2008; Chilingarian \& Mamon 2008; Mieske et al.\ 2004, 2007; Wehner \& Harris 2007; 
Blakeslee \& Barber DeGraaff 2008). However, Romanowsky et al.\ (2009) report the presence of UCDs in the galaxy group hosting NGC 1407. Most recently, a spectroscopically confirmed UCD in the 
low density environment surrounding the Sombrero Galaxy was discovered by Hau et al.\ (2009). UCDs are difficult to find since they are often mistaken for stars, or simply bright globular clusters,
and their structural parameters can only be derived using the Hubble
Space Telescope. The formation of UCDs and possible links between UCDs and globular clusters (either red or blue) remains a subject of debate.

We analyze high resolution Advanced Camera for Surveys (ACS) data of the core
of the Coma cluster, the richest galaxy cluster in the nearby Universe 
(Colless 2001). The Coma cluster provides the unique opportunity to study
 a very large number of distinct galaxies within a small area of the sky, 
as shown in Figure 1. Coma contains a diverse sample of stellar systems 
ranging from globular clusters to giant  elliptical galaxies. 
The core of this cluster contains two giant elliptical galaxies 
(NGC 4874 and NGC 4889) with globular cluster systems that, when combined, 
have more than 30000 members (Blakeslee \& Tonry 1995; Harris et al.\ 2009). 
We specifically study an ACS pointing containing NGC 4874, one of the two central galaxies of  Coma, and the only central galaxy with ACS imaging. The globular cluster system of NGC 4874 alone  
has an estimated {\it minimum} of 18700 members (Harris et al.\ 2009), the largest published globular cluster system.

 \begin{figure*}
 \plotone{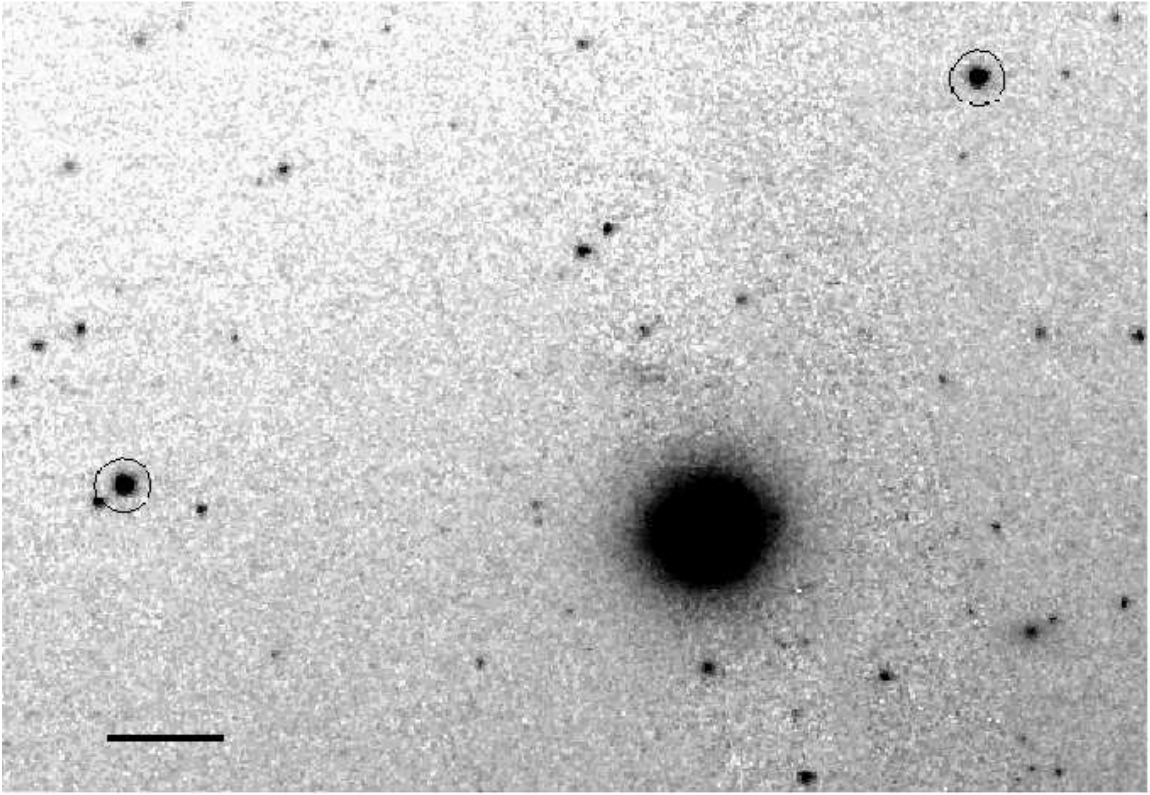} 
 \caption{Section of  the HST/ACS image illustrating the rich diversity 
 of stellar systems in
 the core of the Coma cluster. The brightest galaxy is CcGV19a,
 classified as a compact elliptical by Price et al.\ (2009). Two 
 UCD candidates are indicated with circles. Several members of the rich
 globular cluster system of NGC 4874 are also clearly visible. The
 center of this image is located at $\sim$24 kpc NE from NGC 4874. The
 diffuse halo of this central dominant galaxy is apparent as a diagonal light gradient across  the field of view from bottom right to top left. The bar on the
 lower left is 1 kpc in length. North is up and east is
 left.\label{fig1}} 
 \end{figure*}

 \begin{figure}
 \plotone{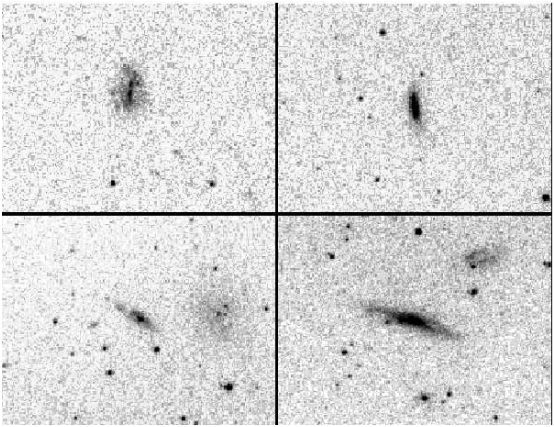} 
 \caption{Four examples of spurious detections eliminated based on their
 suspicious structural parameters. These sources are background galaxies in all likelihood and have effective radii that would correspond to hundreds of parsecs if they were at the distance of Coma. 
\label{fig2}} 
 \end{figure}


We derive a color-magnitude diagram for this immense GCS allowing us to 
clearly define its color distribution. We also estimate the structural parameters of sources that show
extended structure using {\sc ishape}  (Larsen 1999), a 
specially designed software to derive the structural parameters 
of slightly resolved astronomical objects. We find that 52 objects 
have effective radii $\gtrsim$10~pc. These objects have colors, and magnitudes similar
to the brightest globular clusters. We postulate that these objects
constitute the UCD and DGTO population of the core of the Coma Cluster. 

There is a vast terminology to define objects larger than GCs and 
smaller than dwarf elliptical galaxies: Intermediate Mass Object, 
Dwarf Galaxy Transition Object, super-massive star cluster, etc.\ (see Hilker 2006 for a review). UCDs, in particular, are conventionally defined as objects with radii of $\sim$10-100 pc, although 
departures from this size range have been reported (e.g.\ Evstigneeva et al.\ 2008). A commonly quoted lower limit for the mass of UCDs is $2\times10^6$ $M_{\odot}$ (Mieske et al.\ 2008) and their 
magnitude limit is typically $M_V \lesssim-11$ mag (Mieske et al.\ 2006). UCDs also have higher mass-to-light ratios than globular clusters and the presence of dark matter in these stellar systems is 
still under debate (Hasegan et al.\ 2005; Dabringhausen et al.\ 2008). In this paper we will refer to the ensemble of 52 objects with an effective radius $\gtrsim 10$ pc as UCD/DGTOs. The term UCDs 
will be reserved for the subset of 25 objects which are additionally brighter than the luminosity requirement for the UCD label.

Independent methods give estimates of the distance to the Coma cluster
between 84 and 108 Mpc, see Table 1 of Carter et al.\ (2008). We adopt a
fiducial distance of 100 Mpc and thus a scale of 23 pc per 0.05$\arcsec$ 
(one ACS pixel).


\section{Data} 

We use data taken during the Coma Cluster Survey (Carter et al.\ 2008).
This HST Treasury program was able to obtain eighteen pointings of the
core of the Coma Cluster with the ACS Wide Field Channel detector 
before the electronics failure of the instrument. Each field comprises
 imaging in two different filters: F475W (similar to Sloan $g$) 
and F814W (Cousins I) (Mack et al.\ 2003).

The science data from this observing program were prepared through a
dedicated pipeline that is described in detail by Carter et al.\ (2008).
The final images were created using the {\sc pyraf} task {\sc multidrizzle} 
(Koekemoer et al.\ 2002). We use the public data products from the second release of the Coma Cluster Treasury program which provide the best relative 
alignment between the two filters. In the nomenclature of the second 
data release,  we use field number 19 with target name Coma 3-5. 
The exposure times for this field are 2677 s and 1400 s for the 
F475W and F814W filters, respectively.

\section{PSF}

We build the PSF  using the {\sc daophot} package within {\sc pyraf}.  
We perform photometry of candidate stars using the task {\sc phot}, 
the selection of optimal stars is carried out using {\sc pstselect} 
which selects 12 bright, uncrowded, and unsaturated stars across 
the field. The computation of a luminosity weighted PSF is done 
with the task {\sc psf}. {\sc ishape} requires the PSF to be 
oversampled by a factor of ten, that is the pixel size should 
be a tenth of the native pixel size of the science image. 
We create an oversampled PSF with the task {\sc seepsf}, 
and we provide the resulting PSF as input to {\sc ishape}. 
While several options to calculate the PSF for the HST exist, 
it has been shown that with the sequence of  {\sc daophot}  
tasks described above a reliable and accurate model of the PSF core
can be constructed (Spitler et al.\ 2006; Price at al.\ 2009).


\section{Detection}

We median filter the images in both bands with a box size of $41\times41$ pixels and then subtract the result from the original images to obtain a residual free of galaxy light. We thus facilitate the 
detection of the position of point-like sources. A first catalogue of sources is generated running SExtractor (Bertin \& Arnouts 1996) on the image created by adding the median-subtracted 
images of the two filters. Within SExtractor we use a detection threshold of
$3\sigma$ with a minimum of five pixels above the threshold to trigger an 
extraction. A lower detection limit would generate a large
number of spurious detections while more conservative parameters 
would leave globular clusters undetected. 

We perform a first round of careful  visual inspection of all detections and eliminate spurious sources such as background galaxies and objects along the edge of the chip. We use a novel method to 
carry out a secondary visual inspection of our 4976 remaining detections. We take advantage of the short computation time needed to derive an initial set of structural parameters with {\sc ishape}. 
In order to discard spurious sources from our analysis we follow the 
steps described below.  With the initial set of structural parameters 
obtained with {\sc ishape} we create a list of objects with dubious 
values such as very large ellipticity or conspicuously large effective
radii  ($r_h \gtrsim 100$ pc) for visual inspection. By virtue of the high-resolution imaging of the ACS we can determine that most of these objects are  background galaxies, irregular structures,  or 
gradients of surface brightness without link to any real source. Examples of objects flagged and eliminated due to their suspicious structural parameters (i.e.\ large ellipticity or large effective radius) 
are presented in Figure 2. Using this
method we are able to eliminate 170 false detections in a short time. We are left with 4806 sources, most of them globular clusters belonging to the NGC 4874 GCS.


\section{Structural Parameters}


\subsection{Effective radius}

The high resolution of the Hubble Space Telescope has enabled the
measurement of structural parameters of vast numbers of extragalactic
globular clusters. Using WFPC2 data, Larsen et al.\ (2001) derived size
estimates from globular clusters of seventeen nearby, early-type galaxies
and found that the majority of clusters have effective radii of two to
three parsecs. Based on ACS data, and an independent technique, Jord\'an
et al.\ (2005) derived the effective radii of thousands of globular
clusters from 100 early-type galaxies in the Virgo cluster and report a
median GC effective radius of 2.7 pc. Masters et al.\ (2010), studying the
globular clusters around 43  early-type galaxies in the Fornax cluster found
that the median $r_h$ of a GC is 2.9 pc. These results are in remarkable
agreement with size measurements of GCs in spirals (Spitler
et al.\ 2006; Harris et al. 2010) and with the known median effective 
radii of Galactic globular clusters, i.\ e.\ $<r_h>=3.2$ pc 
(Harris et al.\ 1996). At present, no correlation has been found 
between the luminosity (or mass) of globular clusters and their 
effective radius, however their median $r_h$ is $\sim3$pc.


We fit the light profiles of 4806 sources using {\sc ishape}
(Larsen 1999), a  software commonly used to derive the 
structural parameters of barely resolved astronomical objects, 
such as,  extragalactic globular clusters. {\sc ishape} 
convolves the  point-spread function (PSF) with an analytical 
model of the surface brightness profile and searches for the 
best fit to the data by varying the full width half maximum 
(FWHM) of the synthetic profile. The output given by {\sc ishape} 
for each source comprises the effective radius (or equivalently 
half-light radius), the ratio of the minor over major axis, the 
position angle, the signal-to-noise level, and the reduced 
$\chi^2$ (or goodness of fit). 

Within {\sc ishape} the user can 
choose among several analytical models for the light profile that 
will be used during the fitting process, e.g.\ King, S\'ersic, 
Gaussian. We adopt a King profile with a concentration parameter 
of $c=30$; the concentration parameter $c$ is defined as the ratio 
of the tidal radius over the core radius $c = r_t/r_c$ (King 1962, 1966).

The value of $c=30$ accurately represents the average concentration 
parameter for Milky Way, M31, and NGC 5128 globular clusters (Harris et al.\ 1996; Harris 2009a and references therein). Moreover, the concentration parameters for UCDs in Fornax and Virgo 
derived by Evstigneeva et al.\ (2008) are consistent with a choice of $c=30$. In addition, by using c=30 we facilitate comparison with similar studies e.g.\ Blakeslee \& 
Barber DeGraaff (2008). Also, quantitative tests show
that for partially resolved objects the solution for $r_h$ is very 
insensitive to the particular choice of $c$ (Harris et al.\ 2010).

A signal-to-noise ratio of S/N$>$50 is necessary to obtain reliable
measurements of the structural parameters with {\sc ishape}. This
requirement for the S/N was demonstrated through extensive testing 
of {\sc ishape} by Harris (2009a). Of  our initial 4806 sources for 
which we derived structural parameters, 631 sources conform to this S/N requirement simultaneously in both bands. Hereafter, when considering
structural  parameters we will refer only to those 631 objects with
S/N$>$50 in both bands.

At the distance of the Coma cluster, namely 100 Mpc, the effective
radius of a  globular cluster ($\sim$3 pc) is equivalent to six
milliarcseconds, a radius practically irrecoverable in ACS/WFC data
with a pixel size of 0.05$\arcsec$. Using WFPC2 data, Harris et al. 
(2009a) demonstrated that at the distance of Coma objects with $r_h < 6$ 
pc have light profiles indistinguishable from those of stars. As 
expected, the vast majority of clusters for which we attempted 
to obtain reliable structural parameters remain unresolved and 
thus an estimate of their effective radius cannot be made, see Figure 3. However, with  {\sc ishape} we find 52 sources that are positively resolved.


We derive the structural parameters of all sources independently 
in both bands. We minimize systematics between the two measurements 
by strictly following the same set of steps. We also use the same 
set of stars to create both PSFs. In Figure 3 we present the ratio 
$\frac{r_{h F475W}}{r_{h F814W}}$  plotted against $r_{hF814W}$. 
This figure shows how the ratio converges to $\sim$1 at $r_{h F814W}>9$ pc. 
At larger radii ($> 10$ pc) the agreement between the two measurements 
is excellent and reveals a small but real offset. This small but significant 
difference (12\%) in the value of $r_h$ in these two bands can be expected 
due to physical reasons, notably color gradients within globular clusters (Larsen et al.\ 2008). These authors find that $r_h$ can vary by up to 60\% between the measurements in blue (F333W) and 
red (F814W) bands. For our measurements of UCD/DGTOs the median of the ratio $\frac{r_{h F475W}}{r_{h F814W}}$ is $\mu=0.88^{+0.03}_{-0.02}$.


We select as UCD/DGTO candidates only those objects that are resolved in {\it both} bands. The vertical line in Figure 3 shows that all objects with $r_{h F814W} \geqslant 9.2$ pc are consistently 
resolved in both bands. Most of the UCD/DGTO candidates (45 of them) have an effective radius larger than 10 pc in at least one band, the minimum size commonly quoted for UCDs (Mieske et al.\ 
2004). We include in our list of candidates seven objects slightly below this, somehow arbitrary, cut: these seven objects have an $r_h \geqslant$ 9.5 in one band and are within the errors ($\pm 
0.006\arcsec$ or $\pm2.8$ pc, Harris 2009a)  in agreement with the standard size definition of UCDs. Objects resolved in one band but having zero size in the other are rejected as UCD/DGTO 
candidates, see Figure 3.


We also explore the impact of the fitting radius in pixels ({\sc fitrad}) 
that we use as an input of {\sc ishape}. The UCD candidates are barely 
resolved in the ACS images, that is, they have a FWHM of $\sim 3$ pixels. 
{\sc ishape} is limited to a fit radius of a few pixels (Blakeslee \& 
Barber DeGraaff 2008) and the optimal fitting radius is the result of 
a balance between signal and noise, similar to the evolution of the 
S/N ratio as a function of aperture radius in photometric measurements. 
The selection of candidates is carried out using a {\sc fitrad}=6 pixels. 
The consistency of the measurements with {\sc fitrad}=6 pixels is shown 
in Figure 3 where we compare the values of $r_h$ in both bands.

{\sc ishape} yields robust values of effective radius independent of 
the model selected by the user i.e.\ the value of $r_h$ measured with 
a King profile is expected to be similar to $r_h$ obtained using a 
S\'ersic profile. We exploit this property of {\sc ishape} in 
Figure 4. The aim of Figure 4 is twofold: we show the impact on 
the value of $r_h$ measured by {\sc ishape} of i)  two different 
analytical models, ii) two different fitting radii. The values 
of $r_h$ returned by {\sc ishape} should be equivalent for different (but reasonable) analytical models of the surface brightness profile. In fact, a King profile (King 1962, 1966) with a concentration 
parameter of c=30, and a 
S\'ersic model (S\'ersic 1968) with a S\'ersic index of $n=2$ 
are equivalent for our measurements. This is particularly true 
for the smaller UCD candidates as noted by Evstigneeva et al.\ (2008). 
A larger fitting radius, {\sc fitrad}=15 pixels instead of {\sc fitrad}=6 pixels, gives generally a larger $r_h$. Given that a larger fitting radius 
is more appropriate for bigger sources, we report in Table 1 the 
values obtained using  {\sc fitrad}=15 pixels for those sources with 
$r_h > 25$ pc in the initial run with {\sc fitrad}=6 pixels. 


Globular clusters have a median $r_h$ of $\sim$3 pc but their size 
distribution extends up to eight parsecs (Harris 2009a, Masters et al.\ 2010). UCDs studied by Evstigneeva et al.\ (2008) range in $r_h$ between 
4.0 and 93.2 pc while Blakeslee \& Barber DeGraaff (2008) report UCDs 
with $r_h$ ranging from 11.2 to 90.3 pc with a tentative gap between
large globular clusters (or the most compact UCDs) with $r_h < 20$ pc and large 
UCDs with $r_h > 40$ pc. We find objects with $r_h$ almost continuously 
spanning the range from 10 to 40 pc, we find no indication for a break 
in the size distribution of these compact stellar systems.


Metal-poor globular clusters are on average $\sim 25\%$ larger 
than metal-rich ones. This size difference has been well 
documented by several studies (Jord\'an et al.\ 2005; 
Spitler et al.\ 2006, Harris 2009a; Masters et al.\ 2010). 
This size difference has been explained as the result of 
mass segregation combined with the metallicity dependence 
of main sequence stars (Jord\'an 2004). Harris (2009a) 
postulates that the conditions of formation can also be 
responsible for this size difference while Larsen \& Brodie (2003) 
argue that a size-galactocentric distance trend and projection effects 
are the decisive factors. The correlation 
between size and color (or metallicity) observed for globular 
clusters is blurred at higher masses. In our sample blue UCD/DGTO 
candidates have a median effective radius of $12.2^{+1.7}_{-0.5}$ 
pc while the median $r_h$ for red UCD/DGTO candidates is $11.9^{+4.2}_{-0.6}$ pc.


 \begin{figure} 
 \plotone{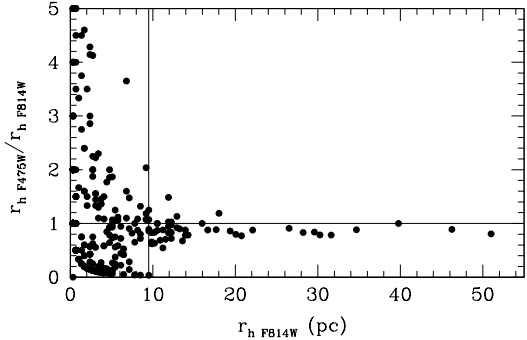} 
 \caption{Ratio of $r_h$ in the two bands vs. $r_h$ in the 
 F814W filter for all systems with S/N$>$50 ({\sc fitrad=6 pixels}). 
This plot reveals how standard globular clusters cannot be 
resolved at the Coma distance. On the other hand, all objects with $r_{h F814W} \geqslant 9.2 $ pc are consistently resolved in both bands as marked by the vertical line. Objects larger than 10 pc 
show an excellent agreement.  \label{fig2}} 
 \end{figure}
 
 \begin{figure} 
 \plotone{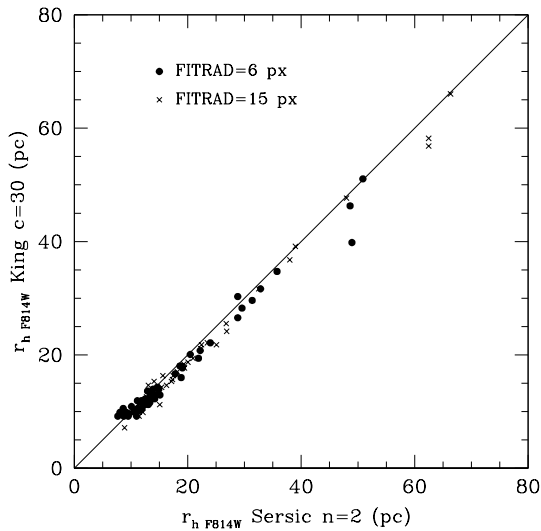} 
 \caption{Effective radii measured by {\sc ishape} using 
different analytical models and different fitting radii 
({\sc fitrad}, in pixels). The effective radius derived by {\sc ishape} 
using different analytical models shows  excellent agreement. 
For the largest UCD candidates a larger fitting radius yields 
a larger effective radius. \label{fig3}} 
 \end{figure}

 \begin{figure} 
 \plotone{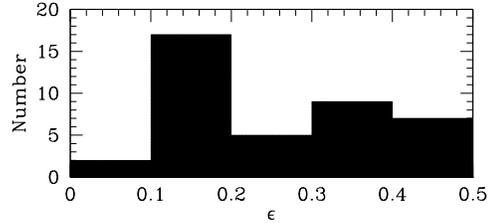} 
 \caption{Ellipticity distribution for UCD/DGTO candidates. A peak in the  
distribution is evident at $\sim 0.15$, and we find only two nearly spherical objects with $\epsilon < 0.1$. \label{fig4}} 
 \end{figure}


\subsection{Ellipticity}

{\sc ishape} computes the minor/major (b/a) axis ratio for each object. 
We use this ratio to derive the ellipticity ($\epsilon = 1- b/a$) and 
to set a selection limit of $0\leq\epsilon<0.5$ for our UCD/DGTO candidates. 
Figure 5 shows the ellipticity values of our UCD/DGTO candidates which span 
the whole range set by our selection criterion. The smallest 
ellipticity of this sample is 0.03, only two candidates have an 
ellipticity of less than 0.1. Blakeslee \& Barber DeGraaff (2008) 
found that all their UCD candidates have an ellipticity between 
0.16 and 0.46. The largest ellipticity in our sample of UCD 
candidates is 0.49 and the median 
of the distribution is 0.24.

The complete list of UCD/DGTO candidates with their effective 
radius and ellipticity is presented in Table 1. Given that {\sc ishape} 
measures the 
effective radius along the major axis, we apply a correction to obtain 
the circularized effective radius that we present in Table 1:

\begin{displaymath}
r_{h, circularized} = r_{h,ishape}\sqrt{(1-\epsilon)}
\end{displaymath}


\section{Photometry}

We perform photometric measurements with the task {\sc phot} of the {\sc
pyraf/daophot} package. We execute aperture photometry with a four pixel
radius and apply aperture correction following the formulas of Sirianni
et al.\ (2005). We use the ACS/WFC Vega zero points for each filter that
we obtain from the updated tables maintained by the STScI, i.e.
$m_{F475W}=26.163$ mag, and  $m_{F814W}=25.520$ mag (Sirianni et al.\ 2006). 
Note that the main catalogue of sources for the Coma Treasury Survey uses 
AB magnitudes (Hammer et al.\ 2010). Foreground Galactic extinction is 
calculated using E(B-V)=0.009 mag (from NED) and the extinction ratios 
of Sirianni et al. (2005), from which we obtain $A_{F475W}= 0.032$ mag 
and $A_{F814W}=0.016$ mag.

We obtain photometric measurements of 4806 sources in F475W and F814W. 
Most of these sources are globular clusters likely belonging to the NGC 4874 
GCS. As shown in Figure 1, crowding is not an issue in our data.


\subsection{Color-Magnitude Diagram}

In the color-magnitude diagram of Figure 6 we also plot the size
information derived with {\sc ishape}. Unresolved globular clusters belonging 
to the NGC 4874 GCS are plotted as black dots. Some of these globular 
clusters might also be intracluster globular clusters of Coma similar to those found in the Virgo cluster (Williams et al.\ 2007; Lee et al.\ 2010). 
UCD/DGTO candidates are plotted as filled red circles. Our 
UCD/DGTO candidates coincide in color with the brightest members of the 
metal-poor and metal-rich subpopulations of globular clusters. 

The color distribution of the brightest GCs is clearly defined 
in the CMD. Background galaxies removed while cleaning spurious detections
have different colors than globular clusters. This was demonstrated by Dirsch et al.\ (2003; their Figure 3) 
studying a field containing the NGC 1399 GCS and a background 
field; see also Harris (2009b) for his study of contamination 
in the M87 field. Moreover, evolutionary synthesis models 
predict that elliptical galaxies of any age, which are the background 
objects most likely to appear as UCDs, have colors $(B-I)>1.9$ 
(Buzzoni 2005).

Mieske et al.\ (2004) set a magnitude limit of $M_V < -11.5$ mag 
for their UCD selection criteria while Hasegan et al.\ (2005)
adopt $M_V < -10.8$ mag. Mieske et al.\ (2006) identified a metallicity break and thus set the onset of the size-luminosity relation at $M_V < -11$ mag. 
Assuming  $V-I = 1.1$ mag and a distance 
modulus to the Coma Cluster of $(m-M)=35.0$ mag we find that 
UCDs should have $m_I<22.9$ mag if we set a selection 
limit of $M_V < -11$ mag. We have assumed $I \approx F814W$ (Sirianni 
et al. 2005). 

There are 110 objects in total with $m_{F814W}<22.9$ mag 
in the color-magnitude diagram of Figure 6, i.e. above the 
horizontal bar, 25 of these objects also have effective radii
characteristic of UCDs. These 110 objects are of similar or higher 
mass than $\omega$ Centauri ($2.5\times10^6 M_{\odot}$, van 
de Ven et al.\ 2006). Higher resolution data should reveal that the remaining 85  objects have an extended structure given that $\sim2\times10^6 M_{\odot}$  marks the onset of a size-luminosity 
relation for stellar structures (Hasegan et al.\ 2005; Forbes et al. 2008; Murray 2009). We also present the color histogram
in the bottom panel of Figure 6, which is consistent with the histogram presented by Harris et al.\ (2009).

\subsection{Size-magnitude}

We attempt to show the onset of the size luminosity relation in Figure 7.
This figure is divided in four quadrants that are limited by the luminosity 
limit of UCDs and the ability of our data to obtain reliable size estimates
i.e.\ a size of $\sim 9.2$ pc. Among the large scatter of the data points characteristic of such a plot (Masters et al. 2010; Forbes et al. 2008) we find that the fraction of unresolved stellar structures 
brighter than the minimum luminosity limit is particularly low around $m_{F814W}\sim 22$ mag, {\it a hint} of the onset of a size-luminosity dependence. More data points are needed to confirm this trend. 


\subsection{Color Bimodality}

We use the software R for our statistical calculations 
presented in this work. R is a freely available, open source 
environment for statistical computing. The code RMIX, run within R, 
is specially designed to evaluate the
mixture of multiple distributions within a histogram. 
RMIX is particularly well suited for the analysis of 
the underlying distributions present in the color-magnitude 
diagram. The code and  its application to the analysis of a 
CMD has been presented, for example, by Wehner et al.\ (2008) 
and Harris (2009a). 

The CMD of globular cluster systems is expected to be bimodal 
and, as noted earlier, for central dominant galaxies this bimodality should
be evident. The bimodal UCD color distribution sits atop a 
bimodal globular cluster distribution well identified for GCs
brighter than F814W $<$ 24 mag, see both panels of Figure 8.

We use the best bimodal fit to the color distribution of 
the 52 UCD/DGTO candidates to estimate the color boundary 
between blue and red UCD candidates (Figure 8). We define 
this point as the intersection of the two Gaussians fitting each 
subpopulation (Peng et al.\ 2006). We thus set the boundary 
between blue and red UCD/DGTO candidates at F475W-F814W = 1.61. 
According to this fit, 63\% of our UCD/DGTOs belong 
to the blue subpopulation, while 37\% are red. For UCDs only (i.e.\ $M_V < -11 $ mag), 44\% of them are blue and 56\% are red.

The brightest candidates are thus red. Self-enrichment models 
developed by Bailin \& Harris (2009) predict  a mass-metallicity 
relationship (MMR) for the most luminous globular clusters 
(with masses $> 10^6 M_{\odot}$). Due to self-enrichment blue
massive globular clusters and UCDs are expected to migrate 
in color towards the red. The visible effect of this 
mass-metallicity relationship in the CMD is the truncation 
towards higher luminosities of the blue subpopulation of globular 
clusters and a red color for the brightest UCDs, as observed in Figure 6
and 8.


 \begin{figure} 
 \plotone{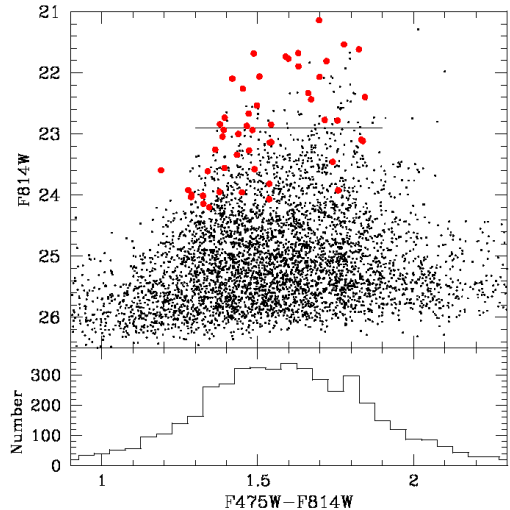} 
 \caption{Top panel: Color-Magnitude Diagram of the globular cluster system and 
 UCD candidates around NGC 4874. Unresolved globular clusters are 
plotted as black dots while resolved UCDs are shown as red circles. 
The horizontal bar marks $M_V =-11$ mag. Bottom panel: Color histogram for all globular clusters. The standard bimodality is blurred due to photometric uncertainties at faint magnitudes. 
\label{fig5}} 
 \end{figure}

 \begin{figure} 
 \plotone{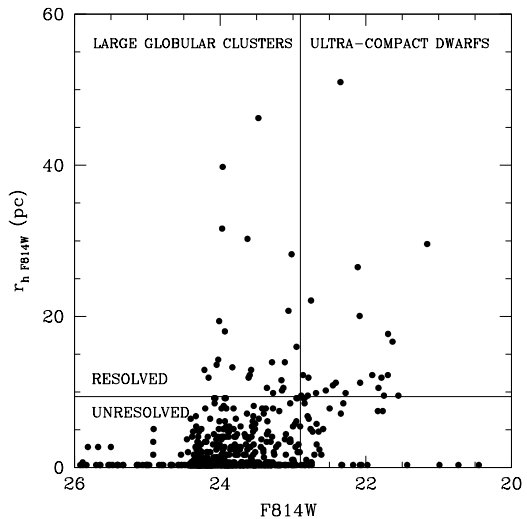} 
 \caption{Magnitude versus effective radius for all objects 
 with S/N $> 50$. This diagram is divided in four quadrants delimited 
 by the minimum luminosity for UCDs and the ability of our data
 to resolve the compact stellar objects we study.\label{fig6}} 
 \end{figure}

 \begin{figure} 
 \plotone{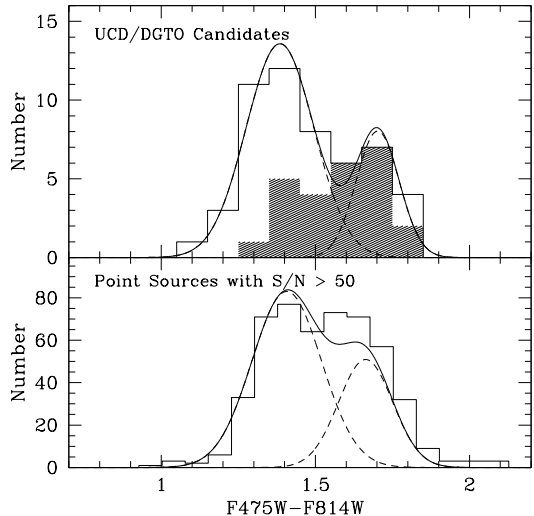} 
 \caption{Top panel: Color distribution of 52 UCD/DGTO candidates. We also over-plot the best bimodal fit to the distribution. Each Gaussian describes 
the red and blue subpopulation. The mean color for the blue subpopulation is $\mu=1.39$ with a dispersion of $\sigma=0.11$ while the mean color for the red subpopulation is $\mu=1.70$ and  $
\sigma=0.07$. The shaded histogram corresponds to 25 UCD candidates, i.e.\ $M_V \lesssim -11$ mag. Bottom panel: color histogram for all 631 sources for which we attempt to derive structural 
parameters, i.e.\ objects with S/N $> 50$. \label{fig6}} 
 \end{figure}


 \begin{figure*} 
 \plotone{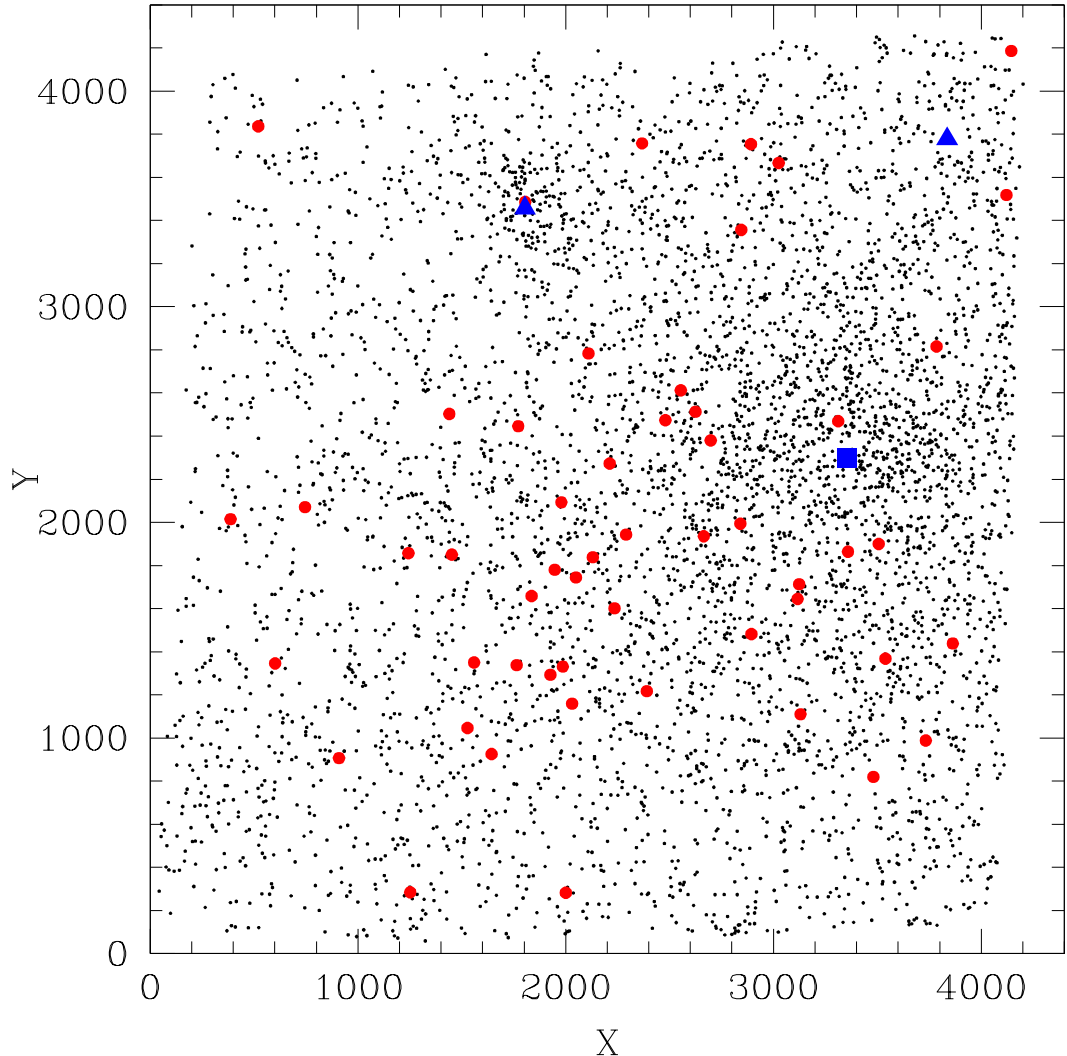} 
 \caption{Spatial distribution of globular clusters (black dots) and UCD/DGTO
candidates (red circles) around NGC 4874 (solid square). An overdensity of 
globular clusters is clearly visible surrounding NGC 4874. A second overdensity 
is associated with NGC 4873  (solid triangle, top middle). NGC 4871 
(triangle, top right) appears to lack a corresponding globular cluster system. 
UCD/DGTO candidates are  anisotropically distributed in the vicinity of NGC 4874. The units are pixels of the ACS/WFC camera. \label{fig7}} 
 \end{figure*}
 
 \begin{figure} 
 \plotone{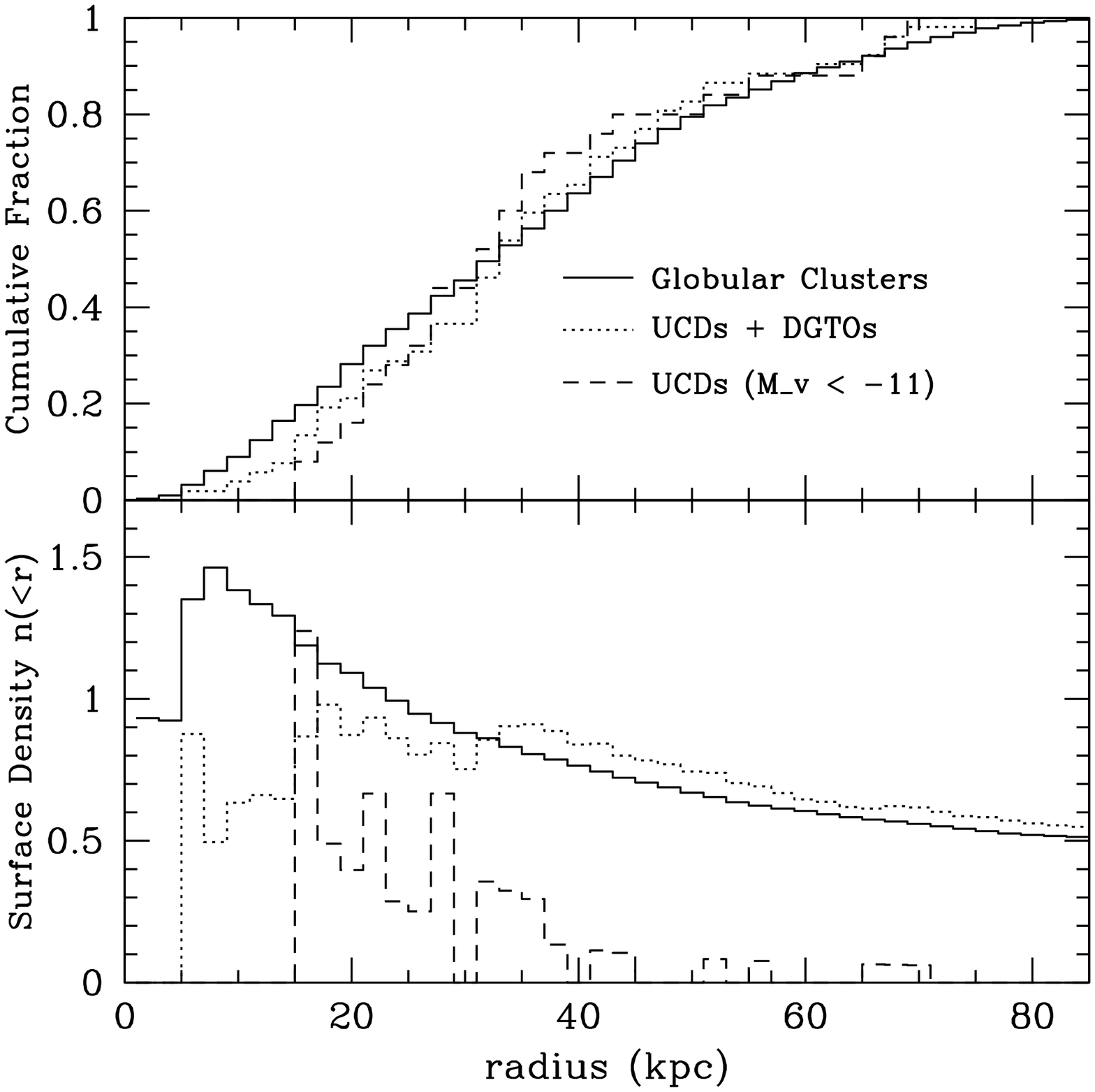} 
 \caption{Top panel: Cumulative fraction of globular clusters, UCD/DGTOs, and UCD candidates as a function of radial distance to NGC 4874. K-S tests on these distributions show that the maximum difference in cumulative fraction between GCs and UCD/DGTOs is $D=0.14$ with $P=0.678$. Between GCs and UCDs these values
 are $D=0.15$ and $P=0.724$. Bottom panel: mean surface density of GCs, UCD/
DGTOs and UCDs. The surface density of UCD/DGTO and UCDs have been scaled by an arbitrary factor of 100 and 500 respectively.\label{fig8}} 
 \end{figure}

\bigskip

\section{Spatial Distribution}

In Figure 9 we plot the spatial distribution of globular clusters, UCD/DGTO
candidates and large galaxies in the ACS frame. At the Coma distance an 
ACS/WFC field of view covers an area of $\sim100$ kpc$^2$. A clear over-density of globular clusters appears to be associated with NGC 4874 
and to a lesser extent with NGC 4873. No over-density is visible around 
NGC 4871. In this image, NGC 4874 is off-center, towards the edge of 
the field of view. The adjacent ACS field was part of the original 
observing program but was not acquired due to the failure of the instrument. A definitive conclusion on the spatial distribution of UCD/DGTO candidates is difficult to reach without the missing adjacent 
field. With the available data we observe that UCD/DGTO candidates appear to congregate around NGC 4874 and more importantly their spatial distribution is not random. The non-uniformity of the 
spatial distribution of UCD/DGTO candidates and their colors argue against background sources masquerading as UCD/DGTOs. 

The cumulative fraction of globular clusters and UCD/DGTO candidates as a function of radial distance to NGC 4874 is plotted in Figure 10. Both stellar systems follow the same distribution. In fact a 
formal two sample 
Kolmogorov-Smirnov test yields a maximum difference between the 
two, azimuthally averaged, cumulative distributions of $D=0.14$ with a corresponding $P=0.678$. This result is valid if we compare UCDs and GCs, the maximum difference in their cumulative fractions is $D=0.15$ with $P=0.724$, see Figure 10. In Figure 10 we also plot the surface density of compact stellar systems in the vicinity of NGC 4874. There is a deficit of UCD/DGTOs in
the inner 15 kpc  around NGC4874. This deficit of DGTOs and UCDs can be explained, at least in part, by tidal interactions with the host galaxy. At small galactocentric distances tidal interactions with the host galaxy destroy or truncate the size of large stellar systems (Hodge 1960, 1962).  For the Milky Way the relation between GC effective radius and galactocentric distance is $r_h\sim R_{gc}^{0.5}$ (van den Bergh et al.\ 1991). Similar relations have been found for several extragalactic GCS  but using projected galactocentric distances (e.g.\ Jord\'an et al. 2005).


\section{Final remarks}

We do not observe any stark discontinuity between the characteristics of 
globular clusters, Dwarf Globular Transition Objects and Ultra-Compact Dwarfs
in the core of the Coma Cluster. On the contrary, bright globular clusters, DGTO,  and UCDs have similar luminosity, color, and effective radius. This is suggestive of a continuous distribution of stellar 
systems gradually increasing in size and luminosity (mass) from globular clusters to UCDs. 

A spectroscopic survey is currently being carried out by Chiboucas 
et al.\ (2009); this project includes several of our UCD candidates as 
targets and will be the next obvious step to confirm the nature 
of our brighter detections.


\acknowledgments

We are grateful to J. Mack (STScI) for her continuous support of the 
ACS data analysis. K. Chiboucas (Gemini) shared with us results of 
her spectroscopic survey of Coma prior to publication, giving us 
greater insight to the data. We received advice from S. Larsen 
(Utrecht) on the optimum {\sc ishape} fitting radius. Many thanks 
to R. Cockcroft (McMaster) for contributing relevant references.
This  research has made use  of the NASA Astrophysics Data System
Bibliographic services (ADS), the NASA/IPAC Extragalactic Database 
(NED),  and  Google. STSDAS and PyRAF are  products  of the  Space
Telescope  Science  Institute, which  is operated by AURA for NASA.
We thank the anonymous referee for a constructive report and a
prompt reply.


{\it Facilities:} \facility{HST (ACS)}


\newpage
\begin{deluxetable}{lcccccccccc}
\tabletypesize{\footnotesize}
\tablecaption{Position, magnitude, color, effective radius, ellipticity, and {\sc fitrad} of UCD/DGTO Candidates \label{tbl-1}} 
\tablewidth{0pt} \tablehead{\colhead{X} & \colhead{Y} & \colhead{RA} &  \colhead{Dec} & \colhead{$m_{F814W}$} & 
\colhead{Color} &\colhead{$r_{h F475W}$} &\colhead{$r_{h F814W}$} & \colhead{$\epsilon$} & \colhead{$r_h, circ$} &
\colhead{\sc fitrad}
}
 
\startdata

    387   &    2015  &   12:59:39.22 & 27:59:54.7 & 19.76 & 1.74 & 41.9 & 47.7 & 0.28 & 40.5 & 15\\
    2390  &    1217  &   12:59:40.60 & 27:58:08.4 & 21.14 & 1.70 & 34.0 & 36.8 & 0.22 & 28.7 & 15\\
    1252  &     284  &   12:59:44.94 & 27:58:54.4 & 21.54 & 1.78 &  8.5 & 9.5  & 0.27 &  8.1 &  6\\
    2894  &    1482  &   12:59:39.22 & 27:57:46.5 & 21.62 & 1.83 & 14.6 & 16.7 & 0.14 & 15.4 &  6\\
    1979  &    2093  &   12:59:37.68 & 27:58:37.6 & 21.68 & 1.63 & 15.6 & 17.7 & 0.34 & 14.4 &  6\\
    1836  &    1658  &   12:59:39.40 & 27:58:40.1 & 21.68 & 1.49 & 12.6 & 12.2 & 0.17 & 11.2 &  6\\     
    3130  &    1110  &   12:59:40.41 & 27:57:31.1 & 21.74 & 1.59 & 11.9 &  9.5 & 0.34 &  7.7 &  6\\
    3481  &     819  &   12:59:41.21 & 27:57:10.9 & 21.77 & 1.60 & 11.2 & 11.9 & 0.20 & 10.6 &  6\\
    2212  &     2273 &   12:59:36.83 & 27:58:28.1 & 21.81 & 1.72 & 10.5 & 10.5 & 0.24 &  9.2 &  6\\
    1764  &    1338  &   12:59:40.64 & 27:58:40.3 & 21.90 & 1.63 &  8.8 & 12.2 & 0.15 & 11.3 &  6\\        
    1527  &    1046  &   12:59:41.91 & 27:58:48.9 & 22.06 & 1.51 &  6.1 & 11.2 & 0.03 & 11.1 &  6\\
    2698  &    2380  &   12:59:36.06 & 27:58:05.4 & 22.07 & 1.70 & 16.0 & 20.1 & 0.18 & 18.2 &  6\\
    2291  &    1943  &   12:59:37.99 & 27:58:20.8 & 22.10 & 1.42 & 24.1 & 26.5 & 0.32 & 21.9 &  6\\
    2891  &    3754  &   12:59:30.83 & 27:58:10.2 & 22.26 & 1.45 &  6.5 & 9.9  & 0.27 &  8.4 &  6\\
    2554  &    2612  &   12:59:35.31 & 27:58:14.9 & 22.33 & 1.66 & 54.1 & 66.0 & 0.04 & 64.7 &  15\\
    2845  &    3357  &   12:59:32.33 & 27:58:08.4 & 22.40 & 1.84 &  9.9 & 11.2 & 0.11 & 10.6 &  6\\
    2001  &     282  &   12:59:44.36 & 27:58:17.8 & 22.44 & 1.67 &  7.5 & 10.9 & 0.35 &  8.8 &  6\\
    2031  &    1159  &   12:59:41.09 & 27:58:25.4 & 22.53 & 1.50 &  8.5 & 10.2 & 0.14 &  9.5 &  6\\    
    2625  &    2513  &   12:59:35.62 & 27:58:10.4 & 22.67 & 1.47 &  8.2 & 9.9  & 0.15 &  9.1 &  6\\
    2235  &    1601  &   12:59:39.30 & 27:58:20.0 & 22.74 & 1.39 & 19.4 & 22.1 & 0.15 & 20.4 &  6\\
    3785  &    2816  &   12:59:33.60 & 27:57:16.8 & 22.77 & 1.72 &  9.9 & 11.9 & 0.20 & 10.6 &  6\\    
     909  &     907  &   12:59:42.90 & 27:59:17.7 & 22.78 & 1.76 &  8.2 & 9.5  & 0.17 &  8.7 &  6\\
    3863  &    1438  &   12:59:38.62 & 27:56:58.7 & 22.84 & 1.38 &  9.9 & 9.2  & 0.27 &  7.8 &  6\\
    3539  &    1368  &   12:59:39.14 & 27:57:13.8 & 22.85 & 1.54 & 11.9 & 12.2 & 0.39 &  9.6 &  6\\
    4121  &    3518  &   12:59:30.74 & 27:57:07.6 & 22.87 & 1.47 & 10.2 & 9.5  & 0.26 &  8.2 &  6\\
    3124  &    1713  &   12:59:38.19 & 27:57:37.7 & 22.94 & 1.48 & 16.0 & 16.0 & 0.32 & 13.2 &  6\\
    1440  &    2503  &   12:59:36.59 & 27:59:08.3 & 22.94 & 1.39 &  9.9 & 9.2  & 0.47 &  6.7 &  6\\   
     602  &    1346  &   12:59:41.52 & 27:59:37.2 & 23.00 & 1.44 & 26.6 & 31.7 & 0.14 & 29.4 & 15\\
     746  &    2071  &   12:59:38.73 & 27:59:37.7 & 23.05 & 1.39 & 16.0 & 20.7 & 0.15 & 19.1 &  6\\
    3507  &    1900  &   12:59:37.20 & 27:57:20.9 & 23.10 & 1.83 & 10.9 & 13.9 & 0.24 & 12.2 &  6\\
    3359  &    1864  &   12:59:37.45 & 27:57:27.7 & 23.12 & 1.84 & 10.5 & 10.5 & 0.46 &  7.7 &  6\\
    3311  &    2469  &   12:59:35.25 & 27:57:36.4 & 23.14 & 1.54 &  6.5 & 10.2 & 0.15 &  9.4 &  6\\
    1559  &    1350  &   12:59:40.76 & 27:58:50.5 & 23.14 & 1.54 &  8.2 & 11.6 & 0.43 &  8.7 &  6\\    
    1644  &     925  &   12:59:42.26 & 27:58:41.9 & 23.26 & 1.36 &  6.1 &  9.9 & 0.43 &  7.4 &  6\\
    2368  &    3758  &   12:59:31.23 & 27:58:35.9 & 23.27 & 1.47 & 12.2 & 13.9 & 0.40 & 10.8 &  6\\
    3025  &    3667  &   12:59:31.05 & 27:58:02.8 & 23.34 & 1.43 &  9.2 & 10.5 & 0.20 &  9.4 &  6\\
     520  &    3837  &   12:59:32.38 & 28:00:07.1 & 23.46 & 1.74 & 55.1 & 56.9 & 0.21 & 50.6 &  15\\
    3733  &     988  &   12:59:40.39 & 27:57:00.4 & 23.56 & 1.39 & 11.9 & 12.9 & 0.38 & 10.2 &  6\\
    2480  &    2474  &   12:59:35.88 & 27:58:17.1 & 23.58 & 1.49 & 10.2 & 12.2 & 0.11 & 11.5 &  6\\    
    2110  &    2783  &   12:59:35.03 & 27:58:38.4 & 23.59 & 1.19 & 17.7 & 11.9 & 0.36 &  9.5 &  6\\
    3116  &    1645  &   12:59:38.44 & 27:57:37.4 & 23.61 & 1.34 & 23.8 & 30.3 & 0.16 & 27.7 &  6\\
    4144  &    4187  &   12:59:28.25 & 27:57:13.4 & 23.82 & 1.54 & 11.9 & 13.3 & 0.24 & 11.6 &  6\\
    1926  &    1293  &   12:59:40.68 & 27:58:31.9 & 23.92 & 1.28 & 21.4 & 18.0 & 0.15 & 16.6 &  6\\
    2049  &    1744  &   12:59:38.92 & 27:58:30.6 & 23.92 & 1.76 & 10.9 &  9.2 & 0.44 & 6.9  &  6\\    
    2665  &    1936  &   12:59:37.72 & 27:58:02.4 & 23.95 & 1.38 & 50.0 & 58.2 & 0.28 & 49.4 &  15\\
    1986  &    1331  &   12:59:40.49 & 27:58:29.4 & 23.96 & 1.45 & 24.8 & 39.2 & 0.38 & 24.9 &  6\\
    2841  &    1994  &   12:59:37.37 & 27:57:54.4 & 24.00 & 1.29 & 16.7 & 19.4 & 0.14 & 18.0 &  6\\
    1772  &    2446  &   12:59:36.54 & 27:58:51.4 & 24.01 & 1.32 & 11.2 & 14.3 & 0.26 & 12.3 &  6\\
    1948  &    1780  &   12:59:38.86 & 27:58:35.9 & 24.03 & 1.29 &  9.2 & 13.6 & 0.49 &  9.7 &  6\\
    1243  &    1857  &   12:59:39.13 & 27:59:11.2 & 24.07 & 1.54 & 18.7 & 9.2  & 0.45 &  6.8 &  6\\
    1452  &    1850  &   12:59:38.99 & 27:59:00.9 & 24.16 & 1.33 & 12.2 & 11.9 & 0.12 & 11.2 &  6\\
    2131  &    1838  &   12:59:38.51 & 27:58:27.6 & 24.22 & 1.34 & 14.6 & 12.9 & 0.28 & 11.0 &  6\\

 \enddata

 \tablecomments{Columns 1 \& 2: x and y position in pixels of the ACS Camera; Columns 3 \& 4:
 RA and Dec (J2000); Column 5: magnitude in the F814W band; Column 6: F475W-F814W color; Column 7: $r_h$ in 
F475W in pc; Column 8: $r_h$ in F814W in pc; Column 9: Ellipticity in F814W; Column 10 Circularized effective radius 
$r_h, circ$  (F814W) in pc; Column 11 {\sc fitrad} in pixels used for the given measurement}

\end{deluxetable}


\newpage




\newpage

\begin{thebibliography}{}

\bibitem[Bailin(2009)]{bai09} Bailin, J. \& Harris, W. E. 2009, ApJ, 695, 1082

\bibitem[Blakeslee \& Barber DeGraaff(2008)]{bla08} Blakeslee, J. P. \&
Barber DeGraaff, R. 2008, AJ, 136, 2295

\bibitem[Blakeslee \& Tonry(1995)]{bla95} Blakeslee, J. P. \&
Tonry, J. L. 1995, ApJ, 442, 579

\bibitem[Bertin(1996)]{ber96} Bertin, E., \& Arnouts, S. 1996, A\&AS, 117, 393

\bibitem[Buzzoni(2005)]{buz05} Buzzoni, A. 2005, MNRAS, 361, 725

\bibitem[Carter(2008)]{car08} Carter, D., et al.\ 2008, ApJS, 176, 424

\bibitem[Chiboucas(2009)]{chi09} Chiboucas, K., et al.\ 2009, BAAS, 41, 234

\bibitem[Chilingarian(2008)]{chl10} Chilingarian I. V. \& Mamon, G. A. 2008, MNRAS, 385, L83


\bibitem[Colles(2001)]{col01} Colless, M. 2001, Coma Cluster in
Encyclopedia of Astronomy and Astrophysics, Nature Publishing Group,
Basingstoke, UK

\bibitem[Dabringhausen(2008)]{dab08} Dabringhausen, J., Hilker, M., \& Kroupa, P. 2008, MNRAS, 386, 864

\bibitem[Drinkwater(2000)]{dri00} Drinkwater, M. J. et al.\ 2000, PASA, 17, 227

\bibitem[Drinkwater(2003)]{dri03} Drinkwater, M. J. et al.\ 2003, Nature, 423, 519

\bibitem[Dirsch(2003)]{dir03} Dirsch, B. et al.\ 2003, AJ, 125, 1908	

\bibitem[Evstigneevaetal(2008)]{evs08} Evstigneeva, E. A. et al.\ 2008, AJ, 136, 461

\bibitem[Forbes(2008)]{for08} Forbes, D. A. et al. 2008, MNRAS, 389,
1924

\bibitem[Hammer(2010)]{Ham10} Hammer, D. et al.\ arXiv:1005.3300

\bibitem[Harris(1996)]{Har09} Harris, W. E. 1996, AJ, 112, 1487

\bibitem[Harris(2009)]{Har09} Harris, W. E. 2009a, ApJ, 699, 254 

\bibitem[Harris(2009)]{Har09} Harris, W. E. 2009a, ApJ, 699, 254 

\bibitem[Harris(2009)]{Har09} Harris, W. E. 2009b, ApJ, 703, 939 

\bibitem[Harrisetal(2009)]{Haretal09} Harris, W. E., et al.\  2009, AJ, 137, 3314 

\bibitem[Harris(2010)]{Har10} Harris, W. E., et al.\ 2010, MNRAS, 401, 1965 

\bibitem[Hasegan(2005)]{Has05} Hasegan, M., et al.\ 2005, ApJ, 627, 203 

\bibitem[Hau(2009)]{Hau09} Hau, G. K., et al.\ 2009, MNRAS, 394, L97 

\bibitem[Hilker(1999)]{hil99}Hilker, M., et al.\ 1999, A\&AS, 134, 75

\bibitem[Hilker(2006)]{Hil09} Hilker, M. 2006, UCDs- A mix bag of 
objects, in ESO Astrophysics Symposia: Globular Clusters Guides to Galaxies, Springer, Berlin

\bibitem[Hodge(1960)]{hod60} 
        Hodge, P. W. 1960, ApJ, 131, 351

\bibitem[Hodge(1962)]{hod62} 
        Hodge, P. W. 1962, PASP, 74, 248

\bibitem[Jordan(2004)]{jor04}Jord\'an, A. 2004, ApJ, 613, L117

\bibitem[Jordan(2005)]{jor05}Jord\'an, A., et al. 2005, ApJ, 634, 1002

\bibitem[King(1962)]{kin62}King, I. R. 1962, AJ, 67, 471

\bibitem[King(1966)]{kin66}King, I. R. 1966, AJ, 71, 64

\bibitem[Koekemoer(2002)]{koe02} Koekemoer, A.  M., Fruchter, A.   S.,
Hook, R.  N., \& Hack,W. 2002, in  The 2002 HST Calibration Workshop:
Hubble after the Installation of the ACS and the NICMOS Cooling System,
ed.  S. Arribas, A.  Koekemoer, \& B.  Whitmore, Baltimore, STScI, 339

\bibitem[Larsen(1999)] {lar99} Larsen, S. S. 1999, A\&AS, 139, 393

\bibitem[Larsen \& Brodie(2003)] {lar03} Larsen, S. S. \& Brodie 2003, ApJ, 593, 
263

\bibitem[Larsen et al.(2001)] {lar01} Larsen, S. S., et al.  2001, AJ, 121, 
2974

\bibitem[Larsen et al.(2008)] {lar08} Larsen, S. S., et al.  2008, MNRAS, 383, 
263

\bibitem[Lee(2010)]{lee10} Lee, M. G., et al.\ 2010, Science, 328, 334

\bibitem[Mack(2003)]{mac03} Mack, J., et al.\ 2003, ACS Data Handbook,
Version 2.0, Baltimore, STScI

\bibitem[Masters(2010)]{mar10} Masters, K., et al.\ 2010, ApJ, 715, 1419 


\bibitem[Mieske(2004)]{mie04} Mieske, S., et al.\ 2004, AJ, 128, 1529

\bibitem[Mieske(2006)]{mie06} Mieske, S., et al.\ 2006, AJ, 131, 2442

\bibitem[Mieske(2007)]{mie07} Mieske, S., et al.\ 2007, A\&A, 472, 111

\bibitem[Mieske(2008)]{mie08} Mieske, S., et al.\ 2008, A\&A, 487, 921

\bibitem[Murray(2009)]{mur09} Murray, N.\ 2009, ApJ, 691, 946 

\bibitem[Peng(2006)]{pen06} Peng, E. W., et al.\ 2006, ApJ, 639, 95

\bibitem[Peng(2008)]{pen08} Peng, E. W., et al.\ 2008, ApJ, 703, 42

\bibitem[Phillipps(2001)]{phi01} Phillipps, S. et al. 2001, ApJ, 560, 201

\bibitem[Price(2009)]{pri09} Price, J. et al.\ 2009, MNRAS, 397, 1816

\bibitem[Romanowsky(2009)]{rom09} Romanowsky, A. J. et al.\ 2009, AJ, 137, 4956

\bibitem[Sersic(1968)]{ser68} S\'ersic, J. L. 1968, Atlas de Galaxias Australes 
(C\'ordoba: Observatorio Astron\'omico, Univ. Nac. C\'ordoba)

\bibitem[Sirianni(2005)]{sir05} Sirianni, M. et al.\ 2005, PASP, 117,
1049

\bibitem[Sirianni(2006)]{sir06} Sirianni, M., Gilliland, R., \& Sembach,
K. 2006 Technical Instrument Report ACS 2006-02, Baltimore, STScI

\bibitem[Spitler(2006)]{spi06} Spitler, L. R. et al.\ 2006, AJ, 132, 1593

\bibitem[vandeven(2006)]{ven06} van de Ven G. et al.\ 2006, A\&A, 445, 513

\bibitem[vandeven(2006)]{ven06} van den Bergh, S. et al.\ 1991, ApJ, 375, 594

\bibitem[Wehner(2007)]{weh07} Wehner, E. M. H. \& Harris, W. E. 2007, ApJ, 668, L35

\bibitem[Wehneretal(2008)]{weh08} Wehner, E. M. H., et al.\ 2008, ApJ, 681, 1233

\bibitem[Williams(2007)]{wil07} William, B. F. et al.\ 2007, ApJ, 654, 835


\end{thebibliography}
\end{document}